\documentclass[10.0pt,a4paper]{article}
\pdfoutput=1
\usepackage{jcappub}
\usepackage{graphicx,wasysym}
\usepackage{rotating}
\usepackage{color}
\usepackage{epsfig}
\usepackage{bm}

\newcommand{\be}{\begin{equation}}
\newcommand{\ee}{\end{equation}}
\newcommand{\bea}{\begin{eqnarray}}
\newcommand{\eea}{\end{eqnarray}}

\def\lsim{\mathrel{\raise.3ex\hbox{$<$\kern-.75em\lower1ex\hbox{$\sim$}}}}
\def\gsim{\mathrel{\raise.3ex\hbox{$>$\kern-.75em\lower1ex\hbox{$\sim$}}}}

\def\lsim{\mathrel{\raise.3ex\hbox{$<$\kern-.75em\lower1ex\hbox{$\sim$}}}}
\def\gsim{\mathrel{\raise.3ex\hbox{$>$\kern-.75em\lower1ex\hbox{$\sim$}}}}

\begin{document}

\hspace*{110mm}{\large \tt FERMILAB-14-240-A}

\title{Challenges in Explaining the Galactic Center Gamma-Ray Excess with Millisecond Pulsars}


\author[a]{Ilias Cholis}
\emailAdd{cholis@fnal.gov}

\author[a,b]{Dan Hooper}
\emailAdd{dhooper@fnal.gov}

\author[b]{Tim Linden}
\emailAdd{trlinden@uchicago.edu}

\affiliation[a]{Fermi National Accelerator Laboratory, Center for Particle Astrophysics, Batavia, IL}
\affiliation[b]{University of Chicago, Kavli Institute for Cosmological Physics, Chicago, IL}

\abstract{
Millisecond pulsars have been discussed as a possible source of the gamma-ray excess observed from the region surrounding the Galactic Center. With this in mind, we use the observed population of bright low-mass X-ray binaries to estimate the number of millisecond pulsars in the Inner Galaxy. This calculation suggests that only $\sim$1-5\% of the excess is produced by millisecond pulsars. We also use the luminosity function derived from local measurements of millisecond pulsars, along with the number of point sources resolved by Fermi, to calculate an upper limit for the diffuse emission from such a population. While this limit is compatible with the millisecond pulsar population implied by the number of low-mass X-ray binaries, it strongly excludes the possibility that most of the excess originates from such objects.}


\maketitle

\section{Introduction}
\label{sec:Intro}
An excess of gamma-rays has been observed from the direction surrounding the Galactic Center, with a spectrum and angular distribution that is in good agreement with that predicted from annihilating dark matter particles~\cite{Daylan:2014rsa,Goodenough:2009gk,Hooper:2010mq,Hooper:2011ti,Abazajian:2012pn,Hooper:2013rwa,Gordon:2013vta,Abazajian:2014fta}. More specifically, this signal can be well fit by 31-40 GeV dark matter particles annihilating to $b\bar{b}$ with a cross section of $\sigma v =(1.7-2.3)\times 10^{-26}$ cm$^3$/s (or by somewhat lighter particles annihilating to lighter quarks with a slightly lower cross section)~\cite{Daylan:2014rsa}. The possibility that this signal constitutes the first detection of particle dark matter interactions has received considerable interest, and many dark matter models have been put forth as potentially viable explanations for the observed excess~\cite{Berlin:2014tja,Izaguirre:2014vva,Alves:2014yha,Agrawal:2014una,Ipek:2014gua,Chang:2014lxa,Wang:2014elb,Balazs:2014jla,Huang:2014cla,McDermott:2014rqa,Cheung:2014lqa,Arina:2014yna,Detmold:2014qqa,Han:2014nba,Cline:2014dwa,Basak:2014sza,Berlin:2014pya,Martin:2014sxa,Ghosh:2014pwa,Abdullah:2014lla,Boehm:2014bia,Ko:2014gha,Cerdeno:2014cda,Hardy:2014dea,Boehm:2014hva,Modak:2013jya,Huang:2013apa,Okada:2013bna,Hagiwara:2013qya,Buckley:2013sca,Anchordoqui:2013pta,Buckley:2011mm,Boucenna:2011hy,Marshall:2011mm,Zhu:2011dz,Buckley:2010ve,Logan:2010nw}.

Due to the complex nature of the Galactic Center region, it is non-trivial to definitely rule out an astrophysical origin of this excess. Any such scenario, however, must be able to account for the following observed characteristics of the signal:
\begin{itemize}
\item{The spectral shape of the excess is measured to strongly peak at energies of $\sim$1-3 GeV (in $E^2 dN/dE$ units). Although previous studies found it difficult to robustly determine the shape of this signal's spectrum at energies below $\sim$1 GeV, the application of cuts to the Fermi event parameter CTBCORE, as applied in Ref.~\cite{Daylan:2014rsa}, have considerably reduced the systematic uncertainties involved in this measurement (see also Ref.~\cite{Portillo:2014ena}). Furthermore, the spectral shape of the excess shows no indication of varying with direction on the sky; the morphological parameters favored by the fit are consistent across all energy bins above 600 MeV~\cite{Daylan:2014rsa}.}
\item{The angular distribution of the excess is approximately spherically symmetric about the Galactic Center. More specifically, the center of the excess is constrained to lie within $\sim$$0.03^{\circ}$ from the Galactic Center (Sgr A$^*$), corresponding to a distance of $\sim$5 parsecs. Any extension of the excess along or perpendicular to the Galactic Plane with an axis ratio greater than $\sim$20\% is also strongly disfavored by the data.}
\end{itemize}

Proposed astrophysical explanations for the gamma-ray excess fall into two categories. The first of these are scenarios in which a $\sim 10^{52}$ erg burst of cosmic rays was injected into the Galactic Center in the recent past ($\sim 10^6$ years ago). Such emission could be dominated by a hadronic and/or leptonic outburst. In the case that the cosmic-ray population is dominated by protons~\cite{Carlson:2014cwa}, the highly aspherical and disk-like distribution of gas leads to a gamma-ray signal that is much less spherically symmetric and much more disk-like than is observed (see also Refs.~\cite{Linden:2012iv,Linden:2012bp}).\footnote{This conclusion can be reached simply by comparing the lower frames of Fig.~3 in Ref.~\cite{Carlson:2014cwa} to the morphology of the excess reported in Ref.~\cite{Daylan:2014rsa}. To address this question more quantitatively, we re-performed the Galactic Center analysis as described in Ref.~\cite{Daylan:2014rsa},  including a proton-burst spatial template (2 Myr or 100 kyr, which were each provided to us by the authors of Ref.~\cite{Carlson:2014cwa}) in place of the spherical dark matter-like template. Our fit found these cosmic ray templates to be incapable of accounting for  any significant amount of the observed emission, and are each disfavored relative to the best fit spherical template at a level of approximately 17$\sigma$.} Furthermore, the spectrum of the excess as reported in Ref.~\cite{Daylan:2014rsa} can only be generated if the cosmic protons are injected with an unrealistic, nearly delta-function-like, spectrum, peaking at $E_{\gamma} \simeq 20-30$ GeV (the more realistic broken power-law models considered in Ref.~\cite{Carlson:2014cwa} do not yield spectra that are compatible with the observed emission)~\cite{Hooper:2010mq,Hooper:2011ti,Hooper:2013rwa}. In the case of a burst dominated by high-energy cosmic ray electrons, in contrast, such an event could potentially yield a somewhat more spherically symmetric distribution of gamma-rays (due to their inverse Compton scattering with radiation rather than with the disk-like distribution of gas)~\cite{Petrovic:2014uda}, although the accompanying bremsstrahlung emission would be disk-like. It is very difficult, however, to simultaneously account for the observed spectrum and morphology of the gamma-ray excess in such a scenario. Furthermore, the energy-dependance of diffusion would lead to a more spatially extended distribution at higher energies, in contrast to the energy-indepenent morphology reported in Ref.~\cite{Daylan:2014rsa}.\footnote{When considering models which invoke extreme physical conditions to account for the excess at the Galactic Center, it may be necessary to reevaluate the contributions from pion production, bremsstrahlung, and inverse Compton emission. In the forthcoming study of Calore {\it et al}.~\cite{Calore:2014xka}, a wide range of diffuse emission models are considered, accounting for a wide variety of physical conditions in the inner region of the Galaxy, finding that a spherical excess with a profile similar to that predicted by dark matter annihilations is preferred by the data in all models (see also Ref.~\cite{Zhou:2014lva}).}

The second category of proposed astrophysical explanations for the gamma-ray excess are scenarios involving a large population of unresolved gamma-ray sources. Millisecond pulsars (MSPs) are known to exhibit a spectral shape that is similar to that of the observed excess, and have thus received some attention within this context~\cite
{Hooper:2010mq,Abazajian:2010zy,Hooper:2011ti,Abazajian:2012pn,Hooper:2013rwa,Gordon:2013vta,Abazajian:2014fta}.  In this letter, we discuss what is known about the spectrum, luminosity function, and spatial distribution of millisecond pulsars in the Milky Way, and use this information to evaluate whether they might be able to account for the observed gamma-ray excess.

\section{The Measured Spectra of Millisecond Pulsars}
\label{sec:spectra}

\begin{figure}[!t]
\hspace{1.0cm}
\begin{center}
\includegraphics[width=3.87 in,angle=0]{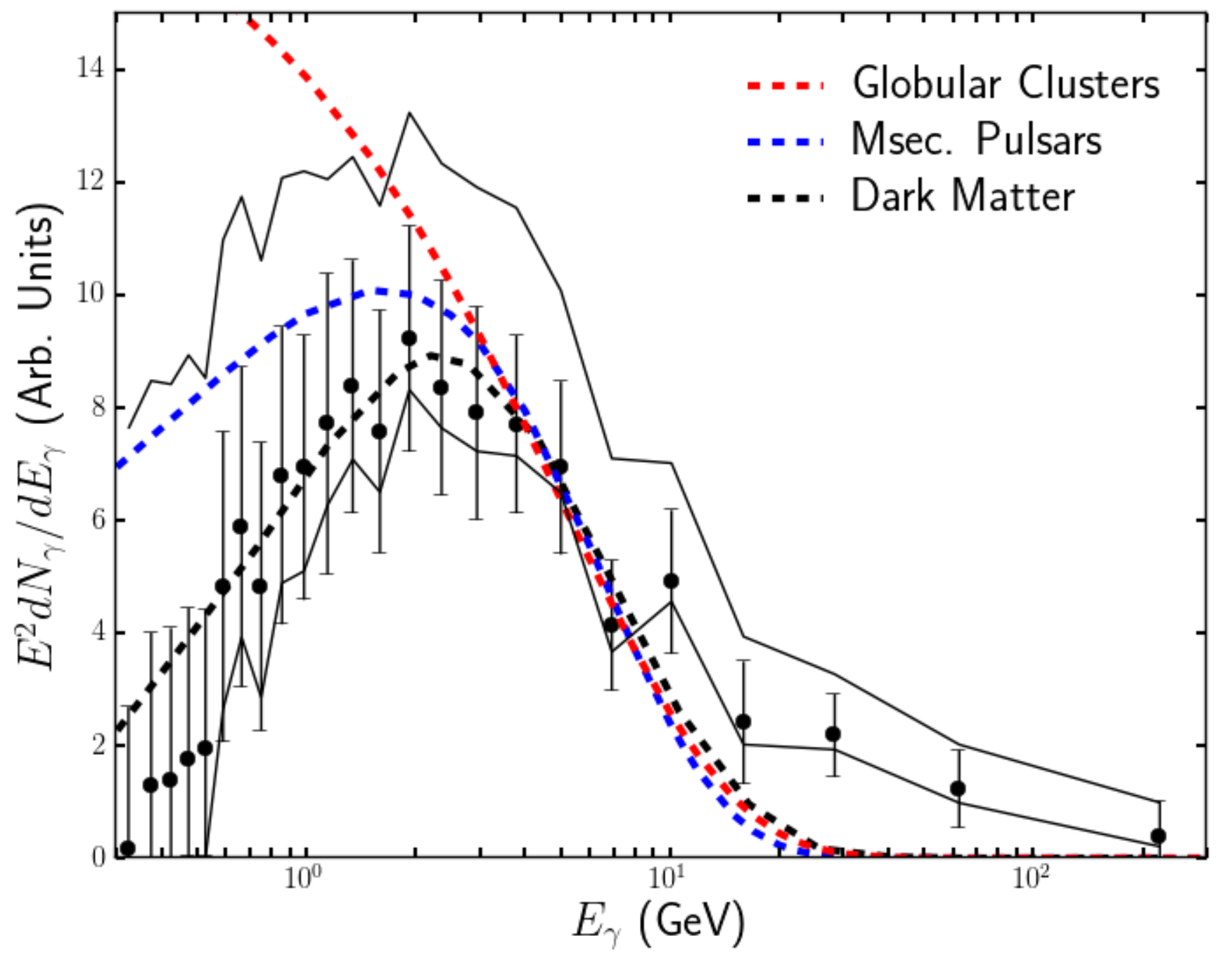}
\caption{The measured spectral shape best fit parameterizaation of the stacked emission from 61 millisecond pulsars observed by Fermi~\cite{Cholis:2014noa} (blue dashed) compared to that of the observed gamma-ray excess correlated systematic errors and envelope galactic center emission (black solid) of~\cite{Calore:2014xka}.  Also shown is the spectral shape best fit of the stacked emission from 36 globular clusters (red dashed)~\cite{Cholis:2014noa}, and the spectrum predicted from a 49 GeV WIMP annihilating to $b{\bar b}$ (black dashed).}

\label{spectra}
\end{center}
\end{figure}

We have recently reported measurements of the gamma-ray spectra of 61 MSPs observed by the Fermi Gamma-Ray Space Telescope, using data collected over a period of 5.6 years~\cite{Cholis:2014noa}. The best-fit spectrum of this collection of (stacked) sources is shown in Fig.~\ref{spectra}, and compared to the spectrum of the observed gamma-ray excess.  Overall, the spectral shape of the gamma-ray excess is fairly similar to that observed from MSPs, and this comparison has motivated an unresolved population of such sources as a possible source of the Galactic Center gamma-ray excess. At energies below $\sim$1 GeV, however, the spectrum observed from MSPs is significantly softer than is exhibited by the excess. 


At this time, a few comments are in order. First, if the observed catalog of gamma-ray MSPs is not representative of the overall population, it is possible that the stacked spectrum could differ from that produced by a large and unbiased collection of such objects. The gamma-ray emission from globular clusters is dominated by MSPs, and their spectra has often been presented as that of an unbiased sample of MSPs.  The spectra observed from Fermi's globular clusters (shown in Fig.~\ref{spectra} as red dashed line~\cite{Cholis:2014noa}) is even softer than that from MSPs~\cite{Cholis:2014noa}, however, and provides a very poor fit to the observed excess. We note that, in addition to the possibility that the very soft $\gamma$-ray spectrum of globular clusters indicates the existence of an additional soft MSP component, it is also possible that this emission component stems from the inclusion of an additional $\gamma$-ray source class intrinsic to globular clusters, a fact which would complicate the comparison of the average globular cluster spectrum to the galactic center excess~(see, however, \citep{collaboration:2010bb}).


While the studies of \cite{Cholis:2014noa} significantly enhance our understanding of the low-energy spectrum of MSPs, the low energy ($<$1~GeV) spectrum of the excess near the galactic center has historically been difficult to constrain. This is due to two key factors: the multiplicity of sources very near the galactic center, and the poor angular resolution of the Fermi-LAT telescope at low $\gamma$-ray energies. In particular, Ref.~\citep{Abazajian:2014fta} showed that the use of different spectral models for the galactic center excess can produce similar fits to the data when the low-energy spectrum of background sources are allowed to fluctuate. 

However, the work of Ref.~\citep{Daylan:2014rsa} and Ref.~\citep{Calore:2014xka} have recently improved the low-energy spectral fitting of the gamma-ray excess through the use of three independent techniques. First, an improved inner-galaxy analysis is employed, following the method of~\citep{Hooper:2013rwa}. The region of interest used in these studies removes emission from bright point sources, masks the region $|b|$~$<$~1$^\circ$ along the galactic plane. 
Additionally, Ref.~\cite{Calore:2014xka} has tested the resiliency of the low-energy spectrum against 60 different background models for the galactic gamma-ray diffusion emission, allowing for the calculation of systematic errors in the $\gamma$-ray excess spectrum.

Second, both the galactic center and inner galaxy analyses of ~\citep{Daylan:2014rsa} employ an enhanced point-spread function at low energies by utilizing cuts on the CTBCORE parameter as detailed in Ref.~\citep{Portillo:2014ena}. Prior to the study of Ref.~\cite{Daylan:2014rsa} and their application of cuts to CTBCORE~\cite{Portillo:2014ena}, significant systematic uncertainties complicated the determination of the low-energy spectrum of the gamma-ray excess (for an illustrative example, see Fig.~10 of Ref.~\cite{Abazajian:2014fta}). After cutting on CTBCORE, however, the shape of the low-energy spectrum is much more robust to variations in analysis procedure.

The galactic center analysis utilizes an iterative fitting algorithm is employed in the galactic center analysis of Ref.~\citep{Daylan:2014rsa}, which allows for the fits from different input source spectra to be compared, and the excess to iteratively converge to the best fit spectrum. In Figure~\ref{spectra}, we compare the average spectrum of MSPs and Globular Clusters determined by Ref. \cite{Cholis:2014noa} to the best fit dark matter spectrum from the inner galaxy analyses of Ref.~\citep{Daylan:2014rsa} and Ref.~\citep{Calore:2014xka}. We clarify that the spectral differences shown are specific to the results of the inner galaxy analysis, and it remains possible that systematic errors may affect the determination of the low-energy gamma-ray spectrum within the inner 1-2$^\circ$ around the galactic center, allowing the spectral fits from MSPs to remain consistent with the excess in this region. Finally we note that the observed best fit stacked millisecond pulsar spectrum of \cite{Cholis:2014noa} of $dN/dE \propto E^{-1.57}$$\cdot exp\{-E/3.78 GeV\}$ gives a $\chi^{2}$ of 36.5 to the Galactic Center spectrum of \cite{Calore:2014xka}  once including also the correlated systematic errors, corresponding to a p-value of 0.02.  

\begin{figure*}[!t]
\includegraphics[width=2.95in,angle=0]{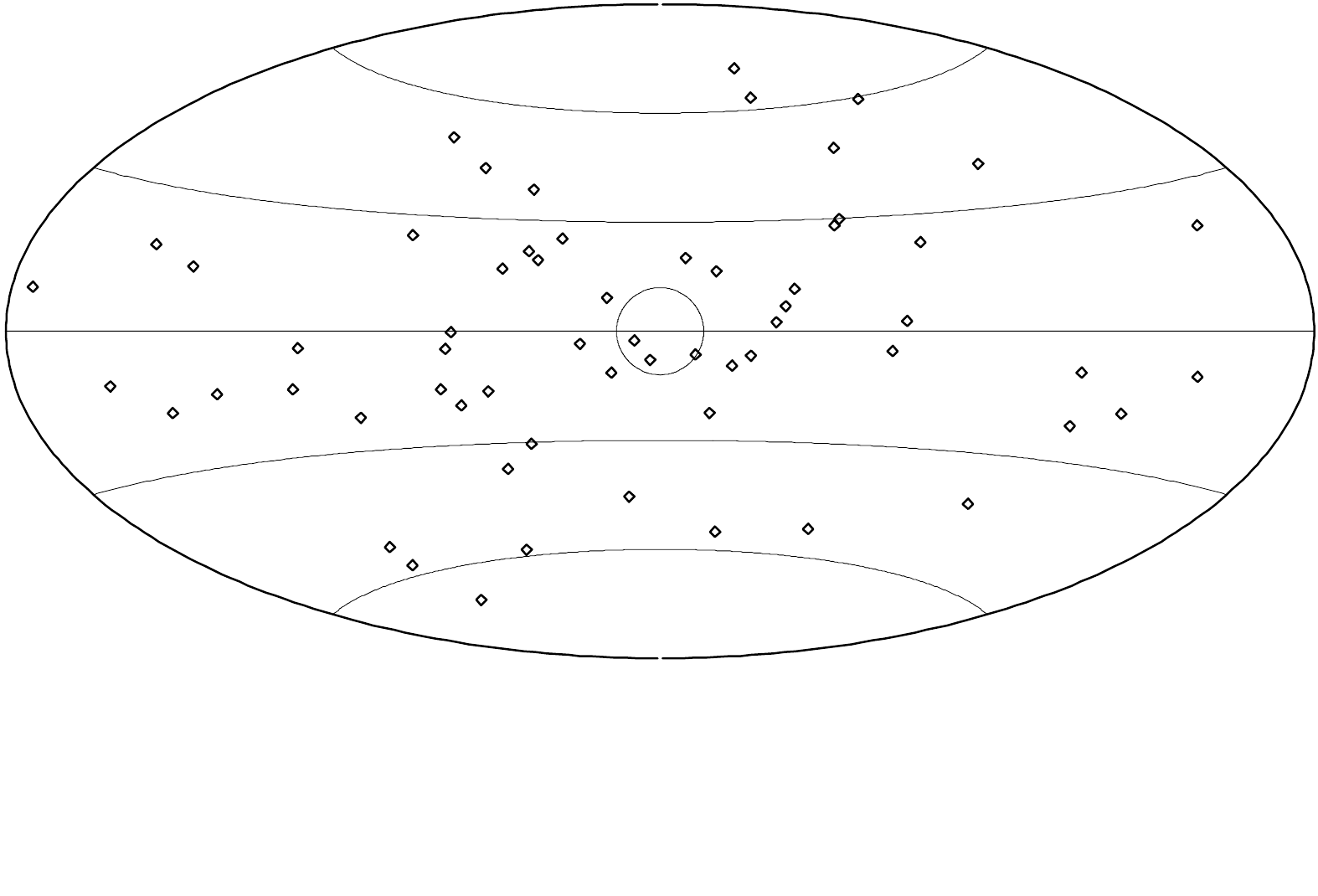}
\hspace{0.3cm}
\includegraphics[width=3.0in,angle=0]{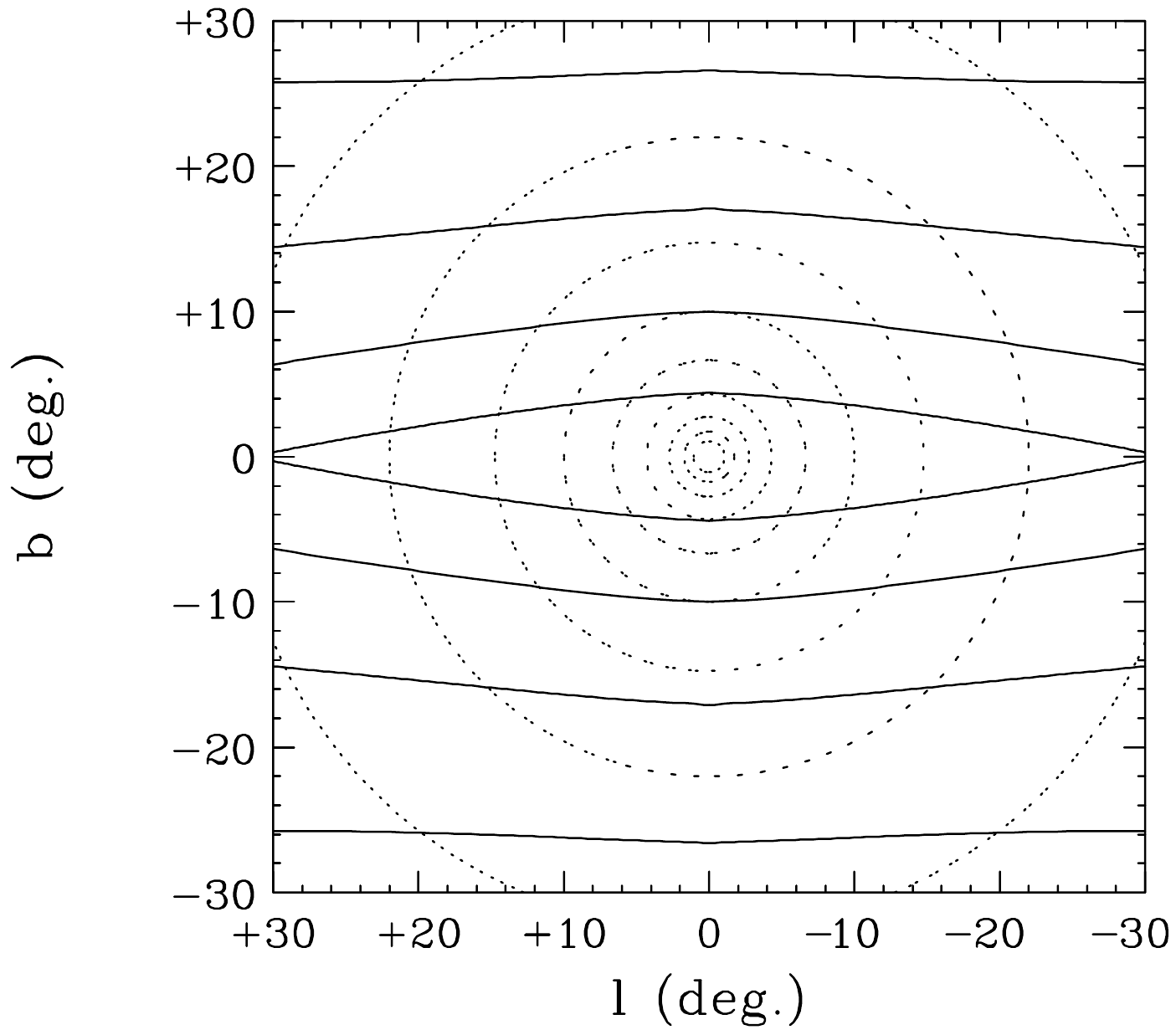}
\caption{In the left frame, we show the location in the sky of the millisecond pulsars currently detected by Fermi, in Aitoff projection. The circle around the Galactic Center represents the $12^{\circ}$ extent to which the gamma-ray excess is currently detected~\cite{Daylan:2014rsa}. In the right frame, we show the morphology of the diffuse gamma-ray emission predicted from millisecond pulsars in the field of the Milky Way (solid) and from annihilating dark matter (dashes).  For millisecond pulsars, we adopt a spatial distribution in cylindrical coordinates given by: $n \propto \exp(-R/5\,{\rm kpc}) \exp(-|z|/1\,{\rm kpc})$, as supported by the catalog of such sources observed by radio and/or gamma-ray wavelengths~\cite{Gregoire:2013yta}. For dark matter, we show here the result for a generalized NFW profile with an inner slope of $\gamma=1.2$. In each case, the lines are contours of constant flux, separated by factors of 2. Given the limits of \emph{Fermi}'s angular resolution, we do not include any contours within the inner 1$^{\circ}$ around the Galactic Center.}
\label{sky}
\end{figure*}

\section{The Observed Distribution of MSPs in the Milky Way}
\label{sec:distribution}
Along with many MSP detections made at radio wavelengths, Fermi has reported the observation of gamma-rays from 62 MSPs. While most of these objects have been found in or around the disk of the Milky Way, some have also been observed to reside within globular clusters.  In the left frame of Fig.~\ref{sky}, we plot the distribution of Fermi's MSPs on the sky. This population has been shown to be well described by a thick disk-like distribution, with an exponential scale height of $\sim$0.5-1.0 kpc~\cite{Hooper:2013nhl,Calore:2014oga}. In the right frame of Fig.~\ref{sky}, we use a MSP thick-disk distribution model fit to this population to estimate the morphology predicted from the unresolved members of this population (solid contours). This prediction is very elongated along the disk, and does not provide a reasonable fit to the much more spherical morphology of the observed excess.  

As it is clear that the MSPs distributed throughout the disk of the Milky Way cannot account for the observed gamma-ray excess, we are instead forced to hypothesize a new (and currently unobserved) population confined to the region surrounding the Galactic Center.  The existence of such a population can be motivated by the fact that the abundance of MSPs (per stellar mass) is much higher in globular clusters than in the disk of the Galaxy. This is generally interpreted as evidence that this MSP population is the result of dynamical interactions, made possible by the high stellar densities found in globular clusters. Given that the number density of stars in the innermost parsec of the Milky Way are comparable to that found in the cores of globular clusters, one expects that a sizable MSP population may be present in the Galactic Center as well. 

\section{Using Low-Mass X-Ray Binaries To Estimate the Number and Distribution of MSPs in the Galactic Center}
\label{sec:LMXBs}
Most MSPs evolved from low-mass X-ray binaries (LMXBs) which consist of a compact object that is powered by accreting matter from a low mass companion. Unlike MSPs, however, the X-ray emission from bright LMXBs can be readily observed in the Inner Galaxy, making it possible to study the distribution of these objects in this region. As different stellar populations of a similar age are expected to contain a similar ratio of MSPs-to-LMXBs, we can use the numbers of LMXBs observed in globular clusters and in the Inner Galaxy to estimate the size of the MSP population in the region surrounding the Galactic Center. 

Focusing on the 16 globular clusters detected as gamma-ray sources by Fermi and reported in Ref.~\cite{Cholis:2014noa}, there are only five ``bright'' ($L > 10^{36}$ erg/s) LMXBs that reside within these systems (in total, 12 bright LMXBs have been detected within all globular clusters).
\footnote{We have adopted a cutoff of $L > 10^{36}$ erg/s in order to ensure that LMXBs would have been observed above the detection threshold of X-Ray instruments in all 16 globular clusters~\cite{White:2001kq, 2000eaa..book.....M} as well as dense region surrounding the galactic center~\cite{Revnivtsev:2008fe, Krivonos:2012sd}. If instead we had adopted a lower threshold, it would not be possible to make a fair comparison of bright LMXBs at globular clusters to bright LXMBs towards the Inner Galaxy. Here and throughout, luminosities denote isotropic equivalent values.} 
The sum of these 16 globular clusters is observed by Fermi to have a total gamma-ray luminosity (above 0.1 GeV) of  $6.1\times 10^{35}$ erg/s, which corresponds to 4.8\% of the luminosity of the Galactic Center excess from within in the innermost $5^{\circ}$. 
If we take these 16 globular clusters to represent a fair sample of both MSPs and LMXBs, we can use the observed gamma-ray emission to calculate how many bright LMXBs should be present within the Inner Galaxy if MSPs are the source of the excess GeV emission.  This calculation finds that if MSPs are to account for the GeV excess, there should also be $103.0^{+69.7}_{-44.5}$ bright LMXBs within 5$^{\circ}$ of the Galactic Center.  
In contrast, INTEGRAL (which has sensitivity in the direction of the Galactic Bulge well beyond the level required to detect such bright sources) has detected only 7 bright LMXB candidates in this region of the sky~\cite{Revnivtsev:2008fe, Krivonos:2012sd}, suggesting that only $\sim$7\% of the GeV excess originates from MSPs. 
This is likely to be an overestimate for two reasons, however. First, by using only the subset of globular clusters detected by Fermi, we have biased our sample towards those systems with especially gamma-ray bright MSPs. Given that the gamma-ray emission from globular clusters is generally dominated by only one or a small number of bright MSPs~\cite{Cholis:2014noa}, the impact of this bias could be significant. 
Second, we note that the stellar populations in globular clusters are generally older than the average stellar population near the Galactic Center. Due to the fact that the LMXB phase precedes the MSP phase and that LMXBs are relatively short lived ($\lsim$10$^8$~yr~\cite{Humphrey:2006et}, compared to $\sim$10$^{10}$~yr for MSPs), we expect the MSP-to-LMXB ratio to be higher in globular clusters than in the Galactic Center. Taken together, this information leads us to estimate that $\sim$1-5\% of the Galactic Center's GeV excess is likely to be the result of unresolved MSPs. 

In addition we note that the LMXBs observed in the Inner Galaxy by INTEGRAL follow a distribution that is very different from that required to produce the morphology of the observed gamma-ray excess. Models of the gamma-ray excess from the galactic center region (10$^\circ$ x 10$^\circ$ ROI), find a morphology which is not extended along the galactic plane by more than 20\% Ref. [1]. However, the observed LMXB population is found to trace the overall stellar population, including a significant degree of elongation along the Galactic Plane~\cite{Revnivtsev:2008fe,Krivonos:2012sd}. Whatever is responsible for the GeV excess is more concentrated around the Galactic Center and is distributed with greater spherical symmetry than is observed among INTEGRAL's LMXB population.   

\section{The Dearth of Gamma-Ray Pulsars Observed in the Inner Galaxy}
\label{sec:LumFunc}
If MSPs are, in fact, responsible for the Galactic Center gamma-ray excess, then Fermi should be able to resolve the brightest of these objects as individual gamma-ray sources~\cite{Hooper:2013nhl}. As a consequence, the number of MSPs (and unidentified gamma-ray sources that could be MSPs) observed in the Inner Galaxy by Fermi can be used to place an upper limit on the total gamma-ray emission from the sum of all MSPs in the region. In Ref.~\cite{Hooper:2013nhl}, this was done using a phenomenological pulsar model, with parameters fit to match the observed MSP population. Here, we instead make use of the MSP luminosity function, as directly determined in Ref.~\cite{Cholis:2014noa}.

In Ref.~\cite{Cholis:2014noa}, we determined the luminosity function of nearby MSPs by studying the sample of such sources detected by Fermi, and correcting for Fermi's distance-dependent luminosity threshold. This measured luminosity function extends down to $L_{\gamma}=10^{31.5}$ erg/s ($E>0.1$ GeV), below which Fermi is unable to resolve any but the most nearby sources. 
For MSPs with a luminosity exceeding this value, their mean luminosity is $9.8\times 10^{33}$ erg/s. In order for such sources to produce the observed intensity of the gamma-ray excess, approximately $2.0 \times 10^3 \times (1-f)$ MSPs with $L_{\gamma} > 10^{31.5}$ erg/s  would be required within the inner 1.8 kpc around the Galactic Center. 
The fraction of the total luminosity from MSPs that comes from sources with $L_{\gamma} < 10^{31.5}$ erg/s was shown in Ref.~\cite{Cholis:2014noa} to be small, $f \ll 1$ (all indications are that the vast majority of the total emission from MSPs comes from a relatively small number of bright sources). 
From the luminosity function and its errors, we calculate that $(226.9^{+91.2}_{-67.4}) \times (1-f)$ of these sources are expected to be bright ($L_{\gamma} > 10^{34}$ erg/s) and $(61.9^{+60.2}_{-33.7}) \times (1-f)$ are expected to be very bright ($L_{\gamma} > 10^{35}$ erg/s). To date, Fermi has detected no MSPs from the inner 1.8 kpc around the Galactic Center (the region of the excess). The three MSPs that appear within this angular region of the sky map shown in the left frame of Fig.~\ref{sky} are each known to reside outside of the inner 1.8 kpc, along a line-of-sight between the Solar System and the Inner Galaxy. 
In this region of the sky, there are also seven unidentified sources in the second Fermi source catalog (2FGL)~\cite{Fermi:2011bm} which do not have IR counterparts in the WISE blazar catalog~\cite{Massaro:2013uoa} (J1830.9-3132, J1820.6-3219, J1730.6-2409, J1748.9-3923, J1813.6-2821, J1717.3-2809, J1727.8-2308).\footnote{We note that the number of unidentified sources detected by Fermi in this region does not represent an excess over that observed along other parts of the Galactic Plane, but is consistent with the average number sources detected per solid angle along the inner disk, $-90^{\circ}<l<90^{\circ}$~\cite{Fermi:2011bm}.} In the outer fraction of the region in question ($|b| \sim 10-12^{\circ}$) Fermi's 2FGL catalog should be approximately complete above $L_{\gamma}>10^{34}$ erg/s. Closer to the Galactic Center, Fermi is less sensitive to point sources, but should still be able to resolve very bright MSPs ($L_{\gamma}>10^{35}$ erg/s)~\cite{Fermi:2011bm,TheFermi-LAT:2013ssa}. If MSPs were responsible for the observed GeV excess, Fermi should have resolved on the order of $10^2$ bright point sources from this region of the sky.  The absence of such sources forces us to conclude that no more than $\sim$$10\%$ of this signal originates from such sources.  While this limit excludes MSPs as the primary source of the observed excess, we note that it is compatible with our estimate based on LMXBs presented earlier in this letter (that $\sim$1-5\% of the excess comes from MSPs). 




                  
\section{Discussion and Conclusions}
\label{sec:remarks}
If instead of adopting the luminosity function as determined in Ref.~\cite{Cholis:2014noa}, we could imagine another hypothetical point source population without the very bright members found among MSPs. For example, we could consider a source population with a luminosity function similar to that presented in Ref.~\cite{Cholis:2014noa}, but truncated above $10^{35}$ erg/s. In that case, we find that a population of $4.4\times 10^{3} \times (1-f)$ sources (with $L_{\gamma} > 10^{31.5}$ erg/s) would be required, of which $374.0^{+201.4}_{-137.9}  \times (1-f)$ would be bright ($L_{\gamma} > 10^{34}$ erg/s). Such a population would again be detectable by Fermi. If we instead truncated our hypothetical population's luminosity function above $10^{34}$ erg/s, a population of $1.4\times 10^{4} \times (1-f)$ ($L_{\gamma} > 10^{31.5}$ erg/s) sources could produce the excess, while plausibly being unresolved by Fermi.

A recent analysis has argued that several of the constraints shown above can be avoided if a significant break is introduced into the MSP $\gamma$-ray luminosity function \cite{Petrovic:2014xra}. This has the capacity to decrease the number of observed $\gamma$-ray MSPs at large distances (such as in the galactic center). This remains a possibility, as the cutoff affects primarily the brightest MSP systems, where there are significant Poisson errors in the MSP source count. However, it is worth noting that broken-power law luminosity models that allow MSPs to explain the entire intensity of the galactic center excess provide worse fits to the luminosity function of the observed galactic MSPs than the simple power-law fit. Specifically, these models predict the existence of 0.6 systems with a luminosity above 10$^{35}$~erg~s$^{-1}$, compared to the three systems which are currently observed (J0218+4232, J0614-3329, and J1311-3430), a discrepancy which is significant at p~=~0.02). Furthermore, it is difficult to understand the introduction of a sharp break in the MSP luminosity function, as physically realistic MSPs likely have include significant $\gamma$-ray beaming, which would smear out any intrinsic break in the luminosity function. It is worth stressing, however, that this depends on the assumption, adopted here, that the population of MSPs near the galactic center are produced with similar distributions to those in the galactic plane and globular clusters.

{\it In Summary}, we find that the population of millisecond pulsars in the Inner Galaxy is likely to be responsible for only a small fraction ($\sim$1-5\%) of the observed GeV excess. This conclusion is supported by the low-mass X-ray binary distribution observed by INTEGRAL, and is consistent with the number of gamma-ray point sources detected by Fermi in this region.  If the gamma-ray excess observed from the region surrounding the Galactic Center is produced by any population of gamma-ray point sources, those sources must be consistently faint (with no significant number of sources brighter than $\sim$$10^{34}$ erg/s), and extremely numerous (tens of thousands of sources within the innermost kpc). The luminosity function of millisecond pulsars, in contrast, is observed to extend to at least $\sim$$2 \times 10^{35}$ erg/s~\cite{Cholis:2014noa}.                   
                  
\smallskip                  
                  
{\it Acknowledgements}: We would like to thank Francesca Calore, Christoph Weniger, and Alex Drlica-Wagner for helpful discussions. This work has been supported by the US Department of Energy. TL is supported by the National Aeronautics and Space Administration through Einstein Postdoctoral Fellowship Award Number PF3-140110.              
                  
\bibliography{pulsarletter3}
\bibliographystyle{apsrev}

\end{document}